# ESTIMATING CORRELATION FROM HIGH, LOW, OPENING AND CLOSING PRICES

By L. C. G. Rogers and Fanyin Zhou

*University of Cambridge and Imperial College London*

In earlier studies, the estimation of the volatility of a stock using information on the daily opening, closing, high and low prices has been developed; the additional information in the high and low prices can be incorporated to produce unbiased (or near-unbiased) estimators with substantially lower variance than the simple open–close estimator. This paper tackles the more difficult task of estimating the correlation of two stocks based on the daily opening, closing, high and low prices of each. If we had access to the high and low values of some linear combination of the two log prices, then we could use the univariate results via polarization, but this is *not* data that is available. The actual problem is more challenging; we present an unbiased estimator which halves the variance.

**1. Introduction.** There is no doubt that volatility is a central concept in the theory and application of quantitative finance. In our simplest models, we treat volatility as a constant of the Black–Scholes paradigm, but we quickly discover that the resulting option pricing formula does not fit reality very well, so we consider variants of the basic model, for example, models where the volatility is allowed to be stochastic in some way. (The enormous literature on GARCH models aims to address similar issues, but cannot be viewed as a variant of Black–Scholes, being as it is a firmly discrete-time theory.) It is not our purpose here to survey this huge field; the reader may consult Ghysels, Harvey and Renault (1996), Shephard (2005) for a survey of (some of) what is known on stochastic volatility. Having chosen a particular model for volatility, the question of estimating it now arises. Again, there is no shortage of papers which propose methods of doing just this; see the survey Broto and Ruiz (2004) for further references. How this estimation is to be carried out depends on the nature of the data available and the model to be estimated. For example, if high-frequency data is available, then we may attempt to estimate volatility through the realized variance of the path. There are several reasons why this is not necessarily a good idea. First, as Alizadeh, Brandt and Diebold (2002) argue, microstructure effects such as bid-ask bounce can significantly bias the estimator upward, though this problem can be obviated to a large extent by a more ingenious choice of estimator; see, for example, Barndorff-Nielsen and Shephard (2004), Zhang, Mykland and Aït-Sahalia (2005). Second, we should expect that the estimates made will not show much intertemporal stability (in view of the well-known profile of intraday trading activity). Indeed, the recent work of Barndorff-Nielsen et al. (2007) confirms this, showing estimates of volatility which vary very substantially from day to day. Third, we have to handle a huge amount of data; while this is not in itself a problem, it is reasonable to ask whether the effort (human and computer) is worth the goal and, indeed, whether the additional effort will actually help toward the goal. Much depends on the intended use, but if we want to price options, or make forecasts, a few months into the future, then we should be using calibration data sampled on a comparable time scale and will require *estimates* of volatility; studies of high-frequency realized volatility are not so much *estimating* volatility as *measuring* it.

In this study, we shall suppose that we are interested in estimating volatility and covariances for the purposes of derivative pricing, derivative hedging and forecasting. For the reasons just outlined, we propose to restrict our attention to *daily* price data, for lack of convincing evidence that high-frequency observation helps to this goal. We shall also discuss only the estimation of *constant* volatilities and covariances; if nothing can be done in this simple situation, then nothing can be done in the more general setting. The strand of the literature that we develop in this paper is that of *range-based estimation of volatility*. The idea of using information on the daily high and low prices, as well as the opening and closing prices, goes back a long way, to Parkinson (1980) and Garman and Klass (1980) at least, with further contributions by Beckers (1983), Ball and Torous (1984), Rogers and Satchell (1991), Kunitomo (1992), Yang and Zhang (2000) and Alizadeh, Brandt and Diebold (2002), among others. However, it is only comparatively recently that attention has been given to range-based estimation of *covariance* between different assets; see, for example, Brunetti and Lildholdt (2002), Brandt and Diebold (2006).

The covariance of assets is important for the computation of the prices of derivatives written on many underlyings, such as basket options; the obvious method of estimation (treating the daily log-returns as i.i.d. multivariate Gaussian variables) produces an unbiased estimator of the covariance matrix. The question we address in this paper is "Can information on daily high and low prices be used to make better (i.e., lower mean squared error) unbiased estimates of the covariance matrix?" The studies

Brandt and Diebold (2006), Brunetti and Lildholdt (2002) work with foreign exchange data, where the availability of data on the cross rates means that one is able to observe highs and lows of *linear combinations* of the log asset prices, allowing one to reduce to existing univariate methodology by polarization. However, such an approach would be impossible if assets were equities, say, since we do not have information on the highs and lows of linear combinations of the log asset prices (unless full tick data is available, but this would be a very different question). For such situations, a completely new approach is required; this is what we undertake in this paper.

In Section 2, we shall, without loss of generality, restrict to the situation of two correlated log-Brownian assets, whose rates of growth we shall assume are both zero. This assumption, used by various authors, is quite innocent if the data is being sampled daily, as the growth rate is negligible in comparison with the fluctuations. We aim to construct an unbiased estimator which is a quadratic function of the high, low and closing (log-)price of the two assets, and which has smallest MSE. For correlation $\rho = -1, 0, 1$, the various moments we require are known in closed form, but for other values of $\rho$, not all were known. [The recent paper Rogers and Shepp (2006) fills in the missing answers.] What we do is to search among linear combinations of quadratic functions of the variables (subject to the constraint that the estimator has no bias if $\rho = -1, 0, 1$) for the estimator that has the smallest MSE when $\rho = 0$. This produces a new estimator whose variance is half that of the obvious estimator based solely on closing prices. We present simulation evidence that this advantage appears to be preserved for other values of $\rho$ and is partly robust to departures from Gaussian returns. The form of the estimator is, moreover, insensitive to errors produced by discrete sampling of the underlying Brownian motions, a problem encountered with some other range-based estimators.

**2. Estimating covariance.** We suppose that the log price processes $X_i(t)$, $i = 1, \ldots, n$, are correlated Brownian motions, that is,

$$E[X_i(s)X_j(t)] = \sigma_{ij} \min\{s, t\}$$

for all $i, j$. We write

$$H_j \equiv \max_{0 \le t \le 1} X_j(t), \qquad L_j \equiv \min_{0 \le t \le 1} X_j(t), \qquad S_j = X_j(1)$$

for the high, low and final log price, respectively, over a fixed time interval which we lose no generality in supposing to be $[0, 1]$. We may also restrict our attention to the case of just two assets since we may estimate the entire correlation matrix if we can handle this case.

To state the main theoretical result of the paper, we shall suppose that $X_1$ and $X_2$ are *standard* Brownian motions, that is, $\sigma_{11} = \sigma_{22} = 1$. (We shall see almost immediately that this restriction is unnecessary.) In this case, the only parameter of the problem to be estimated is the correlation $\rho = \sigma_{12}$ and we obtain the following result.

THEOREM 1. *Among all cross-quadratic functionals (by which we mean a linear combination of the terms $H_1H_2$, $H_1L_2$, $L_1H_2$, $L_1L_2$, $H_1S_2$, $L_1S_2$, $S_1H_2$, $S_1L_2$, $S_1S_2$)*

$$\hat{\rho} \equiv \hat{\rho}(H_1, L_1, S_1, H_2, L_2, S_2)$$

*of the high, low and final log-prices of the two assets which satisfy the unbiasedness condition*

(1) $$E_\rho[\hat{\rho}] = \rho \qquad (\rho = -1, 0, 1),$$

*the one whose variance $E_0[\hat{\rho}^2]$ is minimal when $\rho = 0$ is*

(2) $$\hat{\rho} = \frac{1}{2}S_1S_2 + \frac{1}{2(1-2b)}(H_1 + L_1 - S_1)(H_2 + L_2 - S_2).$$

*The constant $b$ is equal to $2\log 2 - 1 \simeq 0.386294$ and the minimized variance is $E_0[\hat{\rho}^2] = 1/2$.*

REMARK. It is now obvious from Theorem 1, by a simple scaling, that for general $\sigma_{ij}$, the estimator

(3) $$\hat{\sigma}_{12} = \frac{1}{2}S_1S_2 + \frac{1}{2(1-2b)}(H_1 + L_1 - S_1)(H_2 + L_2 - S_2)$$

is unbiased for $\sigma_{12}$ when $\rho = -1, 0, 1$, and when $\rho = 0$, minimizes, variance.

PROOF OF THEOREM 1. The goal is to make an unbiased estimator of $\rho$ by forming linear combinations of the nine possible cross terms, $Z_{HH} = H_1H_2, Z_{HL} = H_1L_2, Z_{LH} = L_1H_2, Z_{LL} = L_1L_2, Z_{HS} = H_1S_2, Z_{LS} = L_1S_2, Z_{SH} = S_1H_2, Z_{SL} = S_1L_2$ and $Z_{SS} = S_1S_2$. Now, the means of these products are known for the cases $\rho = -1, 0, 1$ and the recent paper Rogers and Shepp (2006) establishes that

(4) $$EZ_{HH} = f(\rho)$$
$$\equiv \cos\alpha \int_0^\infty d\nu \frac{\cosh\nu\alpha}{\sinh\nu\pi/2} \tanh\nu\gamma,$$

where $\rho = \sin\alpha$, $\alpha \in (-\pi/2, \pi/2)$ and $2\gamma = \alpha + \pi/2$. Table 1 summarizes the situation. We seek a linear combination $\hat{\rho}$ of the nine cross products with the following properties:

(i) $E_\rho[\hat{\rho}] = \rho$ for $\rho = -1, 0, 1$;

(ii) when $\rho = 0$, the variance of $\hat{\rho}$ is minimal.

In order to find a minimum-variance linear combination, we need to know the covariance of $Z \equiv (Z_{HH}, Z_{HL}, Z_{LH}, Z_{LL}, Z_{HS}, Z_{LS}, Z_{SH}, Z_{SL}, Z_{SS})$ when $\rho = 0$. In this case, the two Brownian motions are independent and the entries of the covariance matrix can be computed from the entries of Table 1. For example, $E_0[Z_{HH} Z_{SL}] = E_1[Z_{HS}] \cdot E_1[Z_{HL}] = -b/2$. Routine but tedious calculations lead to the following covariance matrix:

$$(5) \quad V = \begin{pmatrix} 1 & -b & -b & b^2 & 1/2 & -b/2 & 1/2 & -b/2 & 1/4 \\ -b & 1 & b^2 & -b & 1/2 & -b/2 & -b/2 & 1/2 & 1/4 \\ -b & b^2 & 1 & -b & -b/2 & 1/2 & 1/2 & -b/2 & 1/4 \\ b^2 & -b & -b & 1 & -b/2 & 1/2 & -b/2 & 1/2 & 1/4 \\ 1/2 & 1/2 & -b/2 & -b/2 & 1 & -b & 1/4 & 1/4 & 1/2 \\ -b/2 & -b/2 & 1/2 & 1/2 & -b & 1 & 1/4 & 1/4 & 1/2 \\ 1/2 & -b/2 & 1/2 & -b/2 & 1/4 & 1/4 & 1 & -b & 1/2 \\ -b/2 & 1/2 & -b/2 & 1/2 & 1/4 & 1/4 & -b & 1 & 1/2 \\ 1/4 & 1/4 & 1/4 & 1/4 & 1/2 & 1/2 & 1/2 & 1/2 & 1 \end{pmatrix}.$$

Writing

$$m = (1, -b, -b, 1, 1/2, 1/2, 1/2, 1/2, 1)^T,$$
$$y = (1, -1, -1, 1, 0, 0, 0, 0, 0)^T,$$

our objective now is to choose a 9-vector $w$ of weights to minimize $w \cdot Vw$ subject to the constraints that $w \cdot y = 0$ and $w \cdot m = 1$. This simple optimization problem is easily solved: we find that the solution takes the form

$$(6) \quad w = \alpha V^{-1} m + \beta V^{-1} y,$$

TABLE 1
*Means of the components of Z*

|  | $\rho = -1$ | $\rho = 0$ | $\rho = 1$ | $\rho$ |
|---|---|---|---|---|
| $EZ_{HH}$ | $b$ | $2/\pi$ | $1$ | $f(\rho)$ |
| $EZ_{HL}$ | $-1$ | $-2/\pi$ | $-b$ | $-f(-\rho)$ |
| $EZ_{LH}$ | $-1$ | $-2/\pi$ | $-b$ | $-f(-\rho)$ |
| $EZ_{LL}$ | $b$ | $2/\pi$ | $1$ | $f(\rho)$ |
| $EZ_{HS}$ | $-1/2$ | $0$ | $1/2$ | $\rho/2$ |
| $EZ_{LS}$ | $-1/2$ | $0$ | $1/2$ | $\rho/2$ |
| $EZ_{SH}$ | $-1/2$ | $0$ | $1/2$ | $\rho/2$ |
| $EZ_{SL}$ | $-1/2$ | $0$ | $1/2$ | $\rho/2$ |
| $EZ_{SS}$ | $-1$ | $0$ | $1$ | $\rho$ |

where $\alpha, \beta$ are determined by

$$\begin{pmatrix} m \cdot V^{-1} m & m \cdot V^{-1} y \\ y \cdot V^{-1} m & y \cdot V^{-1} y \end{pmatrix} \begin{pmatrix} \alpha \\ \beta \end{pmatrix} = \begin{pmatrix} 1 \\ 0 \end{pmatrix}. \tag{7}$$

Lengthy but routine calculations lead to the final form (2), as claimed, and the value $E_0[\hat{\rho}^2] = 1/2$ is calculated from the explicit forms of $V$, $m$ and $y$. □

REMARK. (i) It is clear that if we are trying to produce an estimate of the covariance matrix of more than two Brownian motions, estimating each entry by means of (2), then the matrix will be rank 2 and nonnegative definite.

(ii) One problem identified in the earlier literature with estimators based on high and low values occurs when we observe the Brownian motions discretely, at $N$ equally spaced times, say we observe $H^{(N)} \equiv \sup\{X(i/N) : i = 0, \ldots, N\}$ and $L^{(N)} \equiv \inf\{X(i/N) : i = 0, \ldots, N\}$, and these substantially underestimate the supremum and overestimate the infimum. A correction is known to deal with this [see Broadie, Glasserman and Kou (1997)], but we see that as we only ever need to calculate $H + L$, the discretization errors *cancel out* on average because of the observation that $H - H^{(N)}$ and $L^{(N)} - L$ have the same distribution, by symmetry.

(iii) The means in the last five lines in Table 1 are exactly linear in $\rho$, whereas the means in the first four are not. The function $f$ is well approximated by a quadratic; the difference between $f$ and its quadratic approximation (which is exact at $\rho = -1, 0, 1$) is never more than 0.65%. However, if we compute the mean of $\hat{\rho}$, we find

$$\begin{aligned}
\varphi(\rho) &\equiv E_\rho[\hat{\rho}] \\
&= \frac{1}{2}\rho + \frac{1}{2(1-2b)} E_\rho[(H_1 + L_1)(H_2 + L_2) \\
&\qquad\qquad\qquad - S_1(H_2 + L_2) - S_2(H_1 + L_1) + S_1 S_2] \\
&= \frac{1}{2}\rho + \frac{1}{2(1-2b)}[2f(\rho) - 2f(-\rho) - \rho].
\end{aligned}$$

Now, if we simply replace the function $f$ by its quadratic approximation, this expression collapses to $\rho$. In other words, replacing $f$ by its quadratic approximation prevents us from understanding and correcting for the bias in the estimator $\hat{\rho}$.

What we propose to do, therefore, is the following. We suppose that we see data from a run of $N$ days and on day $i$, we compute the value $r_i$ (say) of $\hat{\rho}$. We then take the mean $\bar{r}$ of the $r_i$ and use as our estimator of $\rho$

$$\hat{\rho}_{RZ} \equiv \varphi^{-1}(\bar{r}). \tag{8}$$

TABLE 2
*Simulation results for Brownian motion*

| $\rho$ | $\hat{\rho}_0$ | SD($\hat{\rho}_0$) | $\hat{\rho}_{RZ}$ | SD($\hat{\rho}_{RZ}$) | Variance ratio |
|---|---|---|---|---|---|
| $-0.9$ | $-0.9069$ | 1.367 | $-0.9082$ | 0.8831 | 2.3950 |
| $-0.8$ | $-0.7930$ | 1.290 | $-0.7950$ | 0.8396 | 2.3600 |
| $-0.7$ | $-0.7067$ | 1.239 | $-0.7005$ | 0.8079 | 2.3505 |
| $-0.6$ | $-0.5880$ | 1.157 | $-0.5872$ | 0.7678 | 2.2719 |
| $-0.5$ | $-0.5064$ | 1.137 | $-0.5045$ | 0.7680 | 2.1917 |
| $-0.4$ | $-0.4030$ | 1.075 | $-0.3962$ | 0.7377 | 2.1252 |
| $-0.3$ | $-0.2971$ | 1.060 | $-0.2981$ | 0.7178 | 2.1812 |
| $-0.2$ | $-0.2075$ | 1.019 | $-0.1957$ | 0.7056 | 2.0835 |
| $-0.1$ | $-0.0970$ | 1.003 | $-0.1004$ | 0.7101 | 1.9961 |
| 0.0 | $-0.0038$ | 0.999 | $-0.0011$ | 0.7021 | 2.0285 |
| 0.1 | 0.0992 | 1.010 | 0.0943 | 0.7151 | 1.9942 |
| 0.2 | 0.2083 | 1.014 | 0.2086 | 0.7111 | 2.0331 |
| 0.3 | 0.3051 | 1.042 | 0.3028 | 0.7187 | 2.1032 |
| 0.4 | 0.4089 | 1.096 | 0.4037 | 0.7370 | 2.2128 |
| 0.5 | 0.5013 | 1.124 | 0.5055 | 0.7649 | 2.1611 |
| 0.6 | 0.5967 | 1.159 | 0.6032 | 0.7812 | 2.1994 |
| 0.7 | 0.6913 | 1.190 | 0.6946 | 0.7941 | 2.2468 |
| 0.8 | 0.8062 | 1.309 | 0.7979 | 0.8441 | 2.4057 |
| 0.9 | 0.9012 | 1.344 | 0.9042 | 0.8671 | 2.4038 |

Though the function $\varphi$ is not available in closed form, its numerical values can easily be computed at any desired grid of points in $[-1, 1]$ and then interpolated.

**3. Simulation study.** We have carried out a simulation study of the estimators. For each $\rho = -0.9, -0.8, \ldots, 0.9$, we generated 20,000 paths (of duration 1) of correlated standard Brownian motions, with 500 steps on each path, and for each path, we computed and stored the values of $\hat{\rho}_0 \equiv S_1 S_2$ and $\hat{\rho}_{RZ}$. The results are reported in Table 2. We give the sample means and standard deviations of the two estimators for each value of $\rho$ and we also present the ratio of the sample variance of $\hat{\rho}_0$ over the sample variance of $\hat{\rho}_{RZ}$. We see that this ratio is always at least 2, with the smallest value appearing around $\rho = 0$, where theory predicts the value 2 exactly.

We see that both estimators are close to the true values across the entire range of $\rho$-values chosen, but that $\hat{\rho}_{RZ}$ has at most half the variance of the simple estimator $\hat{\rho}_0$.

To check the robustness of the estimator to model assumptions, we repeated the simulation study using a variance gamma (VG) process instead of Brownian motion, once again with 20,000 paths sampled at 500 points in time. The results are reported in Table 3. Probably the most striking feature is the fact that the estimator $\hat{\rho}_{RZ}$ is now very substantially biased, even for moderately small values of $\rho$. We conclude that the use of this estimator is not advisable if we are not satisfied that the underlying process is Brownian motion. Observe that the bias is always in the direction of underestimating the magnitude of the correlation.

TABLE 3
*Simulation results for VG process*

| $\rho$ | $\hat{\rho}_0$ | $\mathrm{SD}(\hat{\rho}_0)$ | $\hat{\rho}_{RZ}$ | $\mathrm{SD}(\hat{\rho}_{RZ})$ | Variance ratio |
|---|---|---|---|---|---|
| $-0.9$ | $-0.8969$ | 2.0253 | $-0.6847$ | 1.1751 | 2.9705 |
| $-0.8$ | $-0.8094$ | 1.9726 | $-0.6112$ | 1.1094 | 3.1619 |
| $-0.7$ | $-0.6681$ | 1.6592 | $-0.525$ | 0.9746 | 2.8982 |
| $-0.6$ | $-0.6054$ | 1.565 | $-0.4683$ | 0.9070 | 2.9771 |
| $-0.5$ | $-0.5041$ | 1.4674 | $-0.3944$ | 0.8512 | 2.972 |
| $-0.4$ | $-0.3928$ | 1.228 | $-0.3133$ | 0.7264 | 2.8579 |
| $-0.3$ | $-0.3017$ | 1.1538 | $-0.2409$ | 0.6792 | 2.8863 |
| $-0.2$ | $-0.2000$ | 1.0383 | $-0.1637$ | 0.6063 | 2.9331 |
| $-0.1$ | $-0.0854$ | 1.0075 | $-0.0779$ | 0.5759 | 3.0607 |
| 0.0 | $-0.0069$ | 0.9940 | $-0.0029$ | 0.5445 | 3.3326 |
| 0.1 | 0.0967 | 0.9975 | 0.0827 | 0.5694 | 3.0695 |
| 0.2 | 0.2057 | 1.0642 | 0.1660 | 0.6150 | 2.9949 |
| 0.3 | 0.3068 | 1.1338 | 0.2470 | 0.6761 | 2.8119 |
| 0.4 | 0.3891 | 1.2734 | 0.3101 | 0.7514 | 2.8722 |
| 0.5 | 0.4883 | 1.4006 | 0.3870 | 0.8192 | 2.9233 |
| 0.6 | 0.5999 | 1.549 | 0.4701 | 0.9150 | 2.8658 |
| 0.7 | 0.7253 | 1.8293 | 0.5515 | 1.0414 | 3.0855 |
| 0.8 | 0.8042 | 1.9081 | 0.6118 | 1.0988 | 3.0155 |
| 0.9 | 0.8941 | 2.0951 | 0.6807 | 1.2121 | 2.988 |

As a further check of robustness, we performed the same simulation, but using a Brownian motion with drift 0.1. The results are reported in Table 4. This time, the bias of $\hat{\rho}_{RZ}$ is small, but the variance advantage persists.

**4. Empirical study.** In this section, we examine a small data set of stock prices on four stocks: Boeing (BA), GlaxoSmithKline (GSK), General Motors (GM) and Proctor & Gamble (PG). The prices were from the NYSE, for the period 4th February 2002 up to 12th July 2006, a period of 1,118 trading days. The data was from Yahoo Finance. The results are presented in Tables 5 and 6, and in Figure 1. Table 5 presents the point estimates (sample means) of the correlation computed first by the simple open–close estimator and second by the estimator $\hat{\rho}_{RZ}$. Table 6 gives the ratio of the sample variances of the two estimators, the sample variance of $\hat{\rho}_{RZ}$ being expressed as a percentage of the sample variance of $\hat{\rho}_0$. We can see that the point estimators of the correlation are reasonably close, but noticeably different in places; however, inspection of Figure 1 shows that the differences are well within sampling error.

Table 4
*Simulation results for Brownian motion with drift 0.1*

| $\rho$ | $\hat{\rho}_0$ | SD($\hat{\rho}_0$) | $\hat{\rho}_{RZ}$ | SD($\hat{\rho}_{RZ}$) | Variance ratio |
|---|---|---|---|---|---|
| −0.9 | −0.8960 | 1.3634 | −0.8898 | 0.8560 | 2.5372 |
| −0.8 | −0.7842 | 1.2769 | −0.7878 | 0.8267 | 2.3857 |
| −0.7 | −0.6874 | 1.2068 | −0.6917 | 0.7910 | 2.3277 |
| −0.6 | −0.5817 | 1.1604 | −0.5840 | 0.7659 | 2.2953 |
| −0.5 | −0.4851 | 1.1123 | −0.4895 | 0.7482 | 2.21 |
| −0.4 | −0.3953 | 1.099 | −0.3961 | 0.7481 | 2.1582 |
| −0.3 | −0.2868 | 1.0469 | −0.2855 | 0.7196 | 2.1167 |
| −0.2 | −0.1851 | 1.0327 | −0.1929 | 0.7229 | 2.0407 |
| −0.1 | −0.0871 | 1.0087 | −0.0935 | 0.7120 | 2.0074 |
| 0.0 | 0.0143 | 0.9994 | 0.0047 | 0.7050 | 2.0093 |
| 0.1 | 0.1104 | 1.0095 | 0.1091 | 0.7082 | 2.0319 |
| 0.2 | 0.2130 | 1.0575 | 0.208 | 0.7196 | 2.1598 |
| 0.3 | 0.3076 | 1.0599 | 0.3005 | 0.7216 | 2.1572 |
| 0.4 | 0.4088 | 1.0831 | 0.4045 | 0.7359 | 2.166 |
| 0.5 | 0.5118 | 1.135 | 0.5062 | 0.7602 | 2.2291 |
| 0.6 | 0.6241 | 1.2004 | 0.6108 | 0.7827 | 2.3523 |
| 0.7 | 0.7157 | 1.2345 | 0.6987 | 0.7981 | 2.3928 |
| 0.8 | 0.8153 | 1.3177 | 0.8015 | 0.8371 | 2.4777 |
| 0.9 | 0.9199 | 1.3979 | 0.9114 | 0.8937 | 2.4465 |

The sample variance of $\hat{\rho}_{RZ}$ is substantially less than the sample variance of the simple estimator $\hat{\rho}_0$, so we see that for this data, the theoretical advantage of $\hat{\rho}_{RZ}$, namely its lower mean-square error, appears to hold.

**5. Conclusions.** We have presented a new estimator for the correlation of asset prices, based on the information contained in daily high, low, open and close prices. In contrast to other studies, we have *not* supposed that the high and low prices of some linear combination of the log prices is available. While this supposition might be reasonable if the assets were currencies (when the cross rates would provide the required information), it would only be possible in the context of equity if high-frequency data were available. We have found a minimum-variance unbiased estimator quadratic in the variables and have investigated its properties. Simulation experiments showed that the estimator behaved as expected for log-Brownian data, but that the performance on simulated variance gamma data was poor. A small-scale study of prices of equity in major US firms showed that the two estimators agreed to within sampling error and that the sample variance of the new estimator was considerably less. As with range-based estimation of volatility, we conclude that range-based estimation of correlation lacks dependable and decisive advantages over the simpler estimators based only on the open–close prices.

Table 5
*Point estimates of correlation*

| | BA | GSK | GM | PG |
|---|---|---|---|---|
| | Estimated correlation matrix using $\hat{\rho}_0$ | | | |
| BA | 1.0000 | 0.3354 | 0.3294 | 0.3201 |
| GSK | | 1.0000 | 0.2987 | 0.3464 |
| GM | | | 1.0000 | 0.2102 |
| PG | | | | 1.0000 |
| | Estimated correlation matrix using $\hat{\rho}_{RZ}$ | | | |
| BA | 1.0000 | 0.2948 | 0.2925 | 0.2562 |
| GSK | | 1.0000 | 0.2208 | 0.3327 |
| GM | | | 1.0000 | 0.2086 |
| PG | | | | 1.0000 |

TABLE 6
*Ratio of sample variances*

| | Ratio of sample variance of $\hat{\rho}_{RZ}$ to sample variance of $\hat{\rho}_0$ (in %) | | | |
|---|---|---|---|---|
| | **BA** | **GSK** | **GM** | **PG** |
| BA | 92.43 | 55.49 | 45.49 | 60.88 |
| GSK | | 54.74 | 45.90 | 55.09 |
| GM | | | 78.02 | 48.12 |
| PG | | | | 54.97 |

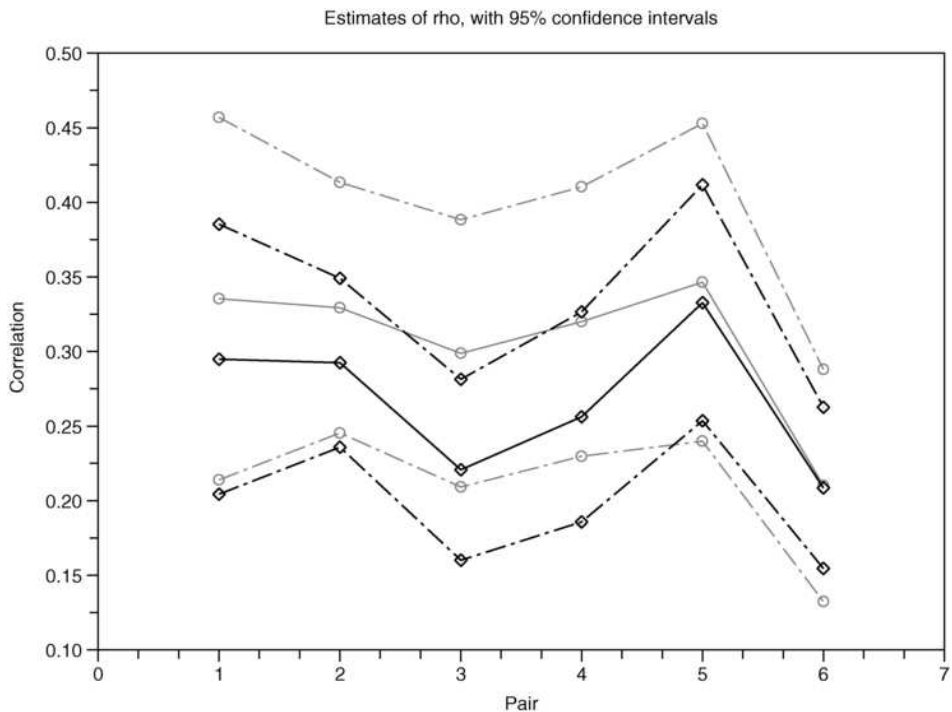

FIG. 1. *Estimates of $\rho$. Estimated values are given by solid lines (circle for simple estimator, diamond for $\hat{\rho}_{RZ}$) and the 95% confidence intervals are given by the dashed lines. The pairs in Figure 1 are listed in the order BA:GSK, BA:GM, GSK:GM, BA:PG, GSK:PG, GM:PG.*

Nevertheless, it seems that it is always worth computing the new estimator, if only as a comparison with the simple open–close estimator. Widely differing numerical values may indicate a departure from log-normality that requires further investigation.

**Acknowledgment.** We thank Nick Brown of BNP Paribas for posing the question which led to this work.

## REFERENCES

ALIZADEH, S., BRANDT, M. and DIEBOLD, F. (2002). Range-based estimation of stochastic volatility models. *J. Finance* **57** 1047–1091.

BALL, C. A. and TOROUS, W. N. (1984). The maximum likelihood estimation of security price volatility: Theory, evidence, and an application to option pricing. *J. Business* **57** 97–112.

BARNDORFF-NIELSEN, O. E., HANSEN, P. R., LUNDE, A. and SHEPHARD, N. (2007). Designing realised kernels to measure the ex-post variation of equity prices in the presence of noise. Technical report, Univ. Oxford.

BARNDORFF-NIELSEN, O. E. and SHEPHARD, N. (2004). Econometric analysis of realized covariation: High frequency based covariance, regression, and correlation in financial economics. *Econometrica* **72** 885–925. MR2051439

BECKERS, S. (1983). Variance of security price returns based on high, low and closing prices. *J. Business* **56** 97–112.

BRANDT, M. W. and DIEBOLD, F. X. (2006). A no-arbitrage approach to range-based estimation of return covariances and correlations. *J. Business* **79** 61–74.

BROADIE, M., GLASSERMAN, P. and KOU, S. G. (1997). A continuity correction for discrete barrier options. *Math. Finance* **7** 325–349. MR1482707

BROTO, C. and RUIZ, E. (2004). Estimation methods for stochastic volatility models. *J. Economic Surveys* **18** 613–649.

BRUNETTI, C. and LILDHOLDT, P. M. (2002). Return-based and range-based (co)variance estimation, with an application to foreign exchange markets. Technical Report 127, Center for Analytical Finance, Univ. Aarhus.

GARMAN, M. and KLASS, M. J. (1980). On the estimation of security price volatilities from historical data. *J. Business* **53** 67–78.

GHYSELS, E., HARVEY, A. C. and RENAULT, E. (1996). Stochastic volatility. In *Statistical Methods in Finance* (G. S. Maddala and C. R. Rao, eds.). *Handbook of Statist.* **14** 119–191. North-Holland, Amsterdam. MR1602124

Kunitomo, N. (1992). Improving the Parkinson method of estimating security price volatilities. *J. Business* **65** 295–302.

Parkinson, M. (1980). The extreme value method for estimating the variance of the rate of return. *J. Business* **53** 61–65.

Rogers, L. C. G. and Satchell, S. E. (1991). Estimating variance from high, low and closing prices. *Ann. Appl. Probab.* **1** 504–512. MR1129771

Rogers, L. C. G. and Shepp, L. (2006). The correlation of the maxima of correlated Brownian motions. *J. Appl. Probab.* **43** 999–999. MR2274808

Shephard, N., ed. (2005). *Stochastic Volatility*: *Selected Readings*. Oxford Univ. Press. MR2203295

Yang, D. and Zhang, Q. (2000). Drift-independent volatility estimation based on high, low, open and close prices. *J. Business* **73** 477–491.

Zhang, L., Mykland, P. A. and Aït-Sahalia, Y. (2005). A tale of two time scales: Determining integrated volatility with noisy high-frequency data. *J. Amer. Statist. Assoc.* **100** 1394–1411. MR2236450

Statistical Laboratory  
University of Cambridge  
Wilberforce Road, Cambridge CB3 0WB  
United Kingdom  
E-mail: l.c.g.rogers@statslab.cam.ac.uk

Institute for Mathematical Sciences  
Imperial College London  
53 Prince's Gate  
South Kensington, London SW7 2PG  
United Kingdom  
E-mail: fanyin.zhou06@imperial.ac.uk